\documentclass[times,twocolumn,final,authoryear]{elsarticle}

\DeclareUnicodeCharacter{0301}{\'{e}}
\usepackage{jasr}
\usepackage{framed,multirow}

\usepackage{amssymb}
\usepackage{latexsym}

\usepackage[switch]{lineno}

\usepackage{url}
\usepackage{xcolor}
\definecolor{newcolor}{rgb}{.8,.349,.1}
\def\revision{}{}

\usepackage[citebordercolor=white]{hyperref}

\journal{Advances in Space Research}

\begin{document}

\verso{M.A. Reiss \textit{et al.}}

\begin{frontmatter}

\title{Unifying the Validation of Ambient Solar Wind Models}

\author[IWF]{Martin~A.~\snm{Reiss}\corref{cor1}}
\cortext[cor1]{Corresponding author: martin.reiss@oeaw.ac.at}

\author[Goddard,CUA]{Karin~\snm{Muglach}}

\author[CCMC]{Richard~\snm{Mullinix}}

\author[CCMC]{Maria~M.~\snm{Kuznetsova}}

\author[CCMC]{Chiu~\snm{Wiegand}}

\author[UniGraz]{Manuela~\snm{Temmer}}

\author[Goddard]{Charles N.~\snm{Arge}}

\author[BuenosAires]{Sergio~\snm{Dasso}}

\author[Goddard]{Shing~F.~\snm{Fung}}

\author[CONACYT]{José Juan~\snm{González-Avilés}}
 
\author[MetOffice]{Siegfried~\snm{Gonzi}}

\author[Goddard]{Lan~\snm{Jian}}

\author[CCMC]{Peter~\snm{MacNeice}}

\author[IWF]{Christian~\snm{Möstl}}

\author[Reading]{Mathew~\snm{Owens}}

\author[KULeuven]{Barbara~\snm{Perri}}

\author[Toulouse, LDE3]{Rui F.~\snm{Pinto}}

\author[CCMC]{Lutz~\snm{Rastätter}}

\author[PredSci]{Pete~\snm{Riley}}

\author[KULeuven]{Evangelia~\snm{Samara}}

\author[]{and \snm{ISWAT~H1-01~Team~Members}}

\address[IWF]{Space Research Institute, Austrian Academy of Sciences, Graz, 8042, Austria}
\address[Goddard]{Heliophysics Science Division, NASA Goddard Space Flight Center, Greenbelt, MD 20771, USA}
\address[CUA]{Catholic University of America, Washington, DC 20064, USA}
\address[CCMC]{Community Coordinated Modeling Center, Code 674, NASA GSFC, Greenbelt, MD 20771, USA}
\address[UniGraz]{University of Graz, Institute of Physics, Graz, 8010, Austria}
\address[BuenosAires]{Instituto de Astronomía y Física del Espacio, UBA-CONICET, Buenos Aires, Argentina}
\address[CONACYT]{Investigadores por M\'exico-CONACYT, Servicio de Clima Espacial M\'exico, Laboratorio Nacional de Clima Espacial, Instituto de Geof\'isca, Unidad Michoac\'an, Universidad Nacional Aut\'onoma de M\'exico, 58190 Morelia, Michoac\'an, M\'exico}
\address[MetOffice]{Met Office, Exeter EX1 3PB, UK}
\address[Reading]{Department of Meteorology, University of Reading, Reading, UK RG6 6BB}
\address[KULeuven]{Centre for Mathematical Plasma Astrophysics, KU Leuven, Celestijnenlaan 200b-box 2400, B-3001 Leuven, Belgium}
\address[Toulouse]{Université de Toulouse; UPS-OMP; IRAP; 31400 Toulouse, France}
\address[LDE3]{LDE3, DAp/AIM, CEA Saclay, 91191 Gif-sur-Yvette, France}
\address[PredSci]{Predictive Science Inc., San Diego, CA 92121, USA}

\received{January 31, 2022}
\accepted{May 16, 2022}

\begin{abstract}
Progress in space weather research and awareness needs community-wide strategies and procedures to evaluate our modeling assets. Here we present the activities of the Ambient Solar Wind Validation Team embedded in the COSPAR ISWAT initiative. We aim to bridge the gap between model developers and end-users to provide the community with an assessment of the state-of-the-art in solar wind forecasting. To this end, we develop an open online platform for validating solar wind models by comparing their solutions with in situ spacecraft measurements. The online platform will allow the space weather community to test the quality of state-of-the-art solar wind models with unified metrics providing an unbiased assessment of progress over time. In this study, we propose a metadata architecture and recommend community-wide forecasting goals and validation metrics. We conclude with a status update of the online platform and outline future perspectives.
\end{abstract}


\end{frontmatter}

\newcommand{\apj}{{\it Astrophys. J.}}
\newcommand{\apjs}{{\it Astrophys. J. Supp.}}
\newcommand{\grl}{{\it Geophys. Res. Lett.}}
\newcommand{\solphys}{{\it Solar Phys.}}


\section{Introduction}
The rate at which we develop and update space weather models has outpaced the rate at which we build our data and validation infrastructure. Consequently, questions such as ''How well does a model perform over a given time interval?'' or ''How much has this model improved over the past five years?'' are, if at all, difficult to answer. Progress in space weather research and awareness, therefore, benefits from community-coordinated strategies and procedures for validation.

Validation is the craft of assessing the quality of models by comparing their solutions with observations. But the validation of space weather models faces challenges. First, keeping up with the ever-growing number of models, different versions thereof, and increasingly versatile user needs without community-wide procedures is practically impossible. Next, the slow, iterative process between model developers and end-users causes a bottleneck. Since new space weather models are usually reported in the literature, models are often outdated when a new paper gets published~\citep[][]{macneice18a}. Finally, adapting to specific user needs is not possible in retrospect in validation studies. End-users have to rely on metrics selected by the authors, which do not always satisfy the increasingly versatile user needs. The \href{https://www.iswat-cospar.org/h1-01}{Ambient Solar Wind Validation Team} embedded in the Committee on Space Research (COSPAR) - \href{https://www.iswat-cospar.org}{International Space Weather Action Teams (ISWAT)} initiative supports the community in these challenges. While the team focuses on validating solar wind models, our strategies can be adapted to adjacent disciplines.

\revision{Assessment of the state-of-the-art in solar wind forecasting is of pivotal importance in space weather research and awareness. The solar wind is a pressure-driven plasma flow that steadily evolves from the solar corona into interplanetary space~\citep{parker58}. Because the solar wind is an excellent electric conductor, the coronal magnetic field is frozen into the solar wind. As the Sun rotates, the solar wind flow drags the coronal magnetic field with it, forming a large-scale structure in the heliosphere that traces out an Archimedean spiral. Fast and slow solar wind streams interacting with each other distort this large-scale structure. We use "ambient solar wind", or simply "solar wind", to refer to this large-scale pattern and exclude interplanetary coronal mass ejections and other solar transients.} 

\revision{\citet{cranmer17} highlighted three reasons why a clear picture of the ambient solar wind is essential.} First, most of the time the solar wind determines the prevailing conditions in our solar system, including bulk speed, and magnetic field strength and orientation~\citep{luhmann02}. Second, fast and slow solar wind flows interacting with each other are a recurrent driver of moderate geomagnetic activity~\citep{kilpua17}, particularly during solar minimum~\citep{verbanac11}. Third, the solar wind sets the conditions through which the most extreme forms of space weather, interplanetary coronal mass ejections, propagate~\citep{gosling99}. Ample evidence shows that solar wind flows can distort and deflect interplanetary coronal mass ejections and thereby affect their severity~\citep{riley97, odstrcil99a, case08, zhou17}.

The validation of solar wind models has received much attention; for instance, \cite{owens05, macneice09a} and \cite{reiss16} developed validation procedures that detect abrupt transitions from slow to fast wind. \cite{owens17} and \cite{henley17} discussed ensemble solar wind forecasts using a "cost-loss" analysis, and \cite{owens18} proposed time window approaches to complement traditional point-to-point comparison metrics. Numerous studies have assessed the quality of ambient solar wind models by comparing their solutions to in situ measurements~\citep[see, for example,][among others]{owens08, macneice09a, norquist10, jian11, gressl14, jian15, jian16, devos14, reiss16, reiss19, hinterreiter19, li20, riley21}. Other observational tests focus on the Earth to Sun magnetic connectivity~\citep{macneice11}, interplanetary scintillation~\citep{kim14, jackson15, gonzi21}, coronagraph images~\citep{jones17, lamy19}, and more. As an unbiased benchmark, \cite{owens13b} and~\cite{kohutova16} proposed persistence models that assume the solar wind condition repeat after each Carrington rotation.

Despite these and other important validation efforts, \revision{the impression among the community is that} model development has greatly outpaced the validation infrastructure development. To rectify this issue, \cite{macneice18b} recommended the development of automated protocols to maintain up-to-date validation results. In line with this recommendation, we formed the Ambient Solar Wind Validation Team with the following aims:
\begin{enumerate}
    \item Develop a comprehensive metadata architecture including metrics to enable sustainable validation of the state-of-the-art and progress assessment over time.
    \item Implement an open online platform in collaboration with NASA's Community Coordinated Modelling Center (CCMC) to validate solar wind models with streamlined metrics.
    \item Quantitatively assesses the state-of-the-art in forecasting the solar wind conditions at Earth and other planetary environments. 
    \item Use our developed infrastructure to establish an "Ambient Solar Wind Scoreboard" to maintain up-to-date validation results.
\end{enumerate}

This study presents our progress concerning Aim~1 and~2, while a follow-up study will focus on Aim~3 and~4. Section~\ref{sec:stateoftheart} reviews the state-of-the-art in the methodology of solar wind forecast models. Section~\ref{sec:metadata} proposes the metadata architecture used to register community models in the online platform. Section~\ref{sec:metrics} recommends community-wide validation metrics, while Section~\ref{sec:platform} presents the current status of the online platform. The discussion in Section~\ref{sec:discussion} concludes this study and outlines our future perspectives. 

\section{Ambient Solar Wind Forecasting in 2022}\label{sec:stateoftheart}

In what follows, we focus on forecast models with possible applications in an operational setting. \revision{Our discussion, therefore, excludes solar wind models that do not provide context for data analysis, and are designed to study physical processes in numerical experiments.} We furthermore limit our discussion to \textit{ambient} solar wind models that exclude the dynamics of interplanetary coronal mass ejections. 

State-of-the-art solar wind frameworks couple models of the global corona and heliosphere. As a standard configuration, the coronal domain spans the distance of 1 solar radius ($\textup{R}_\odot$) to a distance between 20 and 30~$\textup{R}_\odot$ (depending on the model) \revision{beyond the estimated Alfvén Point}, and the heliospheric domain extends between the outer boundary of the coronal domain to 1~AU (or further when needed). 

\begin{figure*}
\centering
\includegraphics[width=0.99\textwidth]{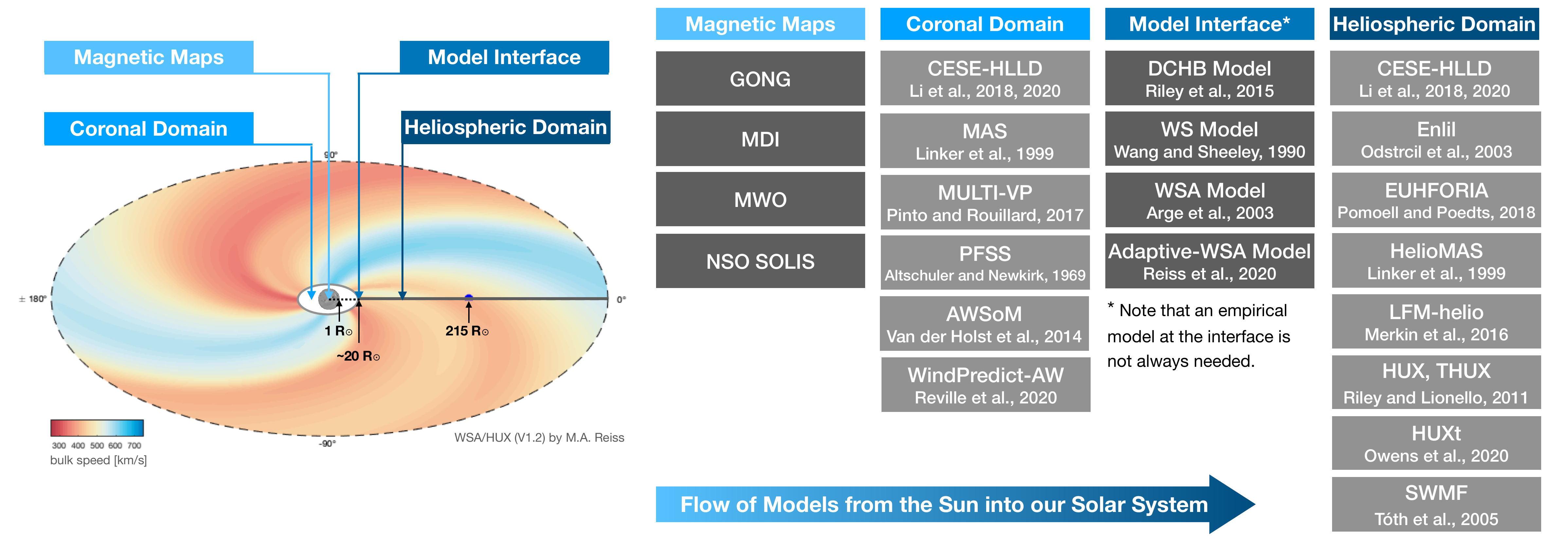}
\caption{Examples of state-of-the-art models for forecasting the solar wind in Earth's space weather environment according to domains.}
\label{fig:stateoftheart}
\end{figure*}

\revision{Figure~\ref{fig:stateoftheart} shows the flow of models starting from the inner boundary of the coronal domain into the heliospheric domain. Synoptic photospheric magnetograms determine the inner boundary in the coronal domain. Ample evidence shows that the choice of synoptic magnetograms has a decisive effect on the solar wind model solutions~\citep[see][]{riley14, pevtsov15, li21, jin22}. Magnetic flux transport models such as the Evolving} Surface Flux Transport Model~\citep[ESFTM;][]{schrijver03}, Air Force Data Assimilative Photospheric Flux Transport~\citep[ADAPT;][]{arge10, hickmann15}, and the SURface Flux Transport~\citep[SURF;][]{upton14} are expected to improve these photospheric magnetic field measurements. ADAPT, for example, creates an ensemble of possible magnetograms to account for differential rotation, meridional flows, and random flux emergence.

In the coronal domain, numerical models rely on extrapolations computed from photospheric magnetograms. One of the workhorse models in the community to reconstruct the global coronal magnetic field is the Potential Field Source Surface model~\citep[PFSS;][]{altschuler69}. The PFSS model assumes that regions of the photosphere are current-free, which means that coronal field solutions can be expressed as the gradient of a scalar potential. Since potential fields give closed field lines, an outer boundary condition is added. This outer boundary is defined by a spherically-symmetric "source surface" where the field is assumed to be only radial (inward or outward-directed). The source surface radius is set to a reference height of 2.5~solar radii to best match observations~\citep{hoeksema82}. Recent studies argue that the source surface radius \revision{in the PFSS model} can vary between 1.5 and 3~solar radii depending on the solar activity cycle~\citep{reville15,asvestari19,boe20}. The Schatten Current Sheet model~\citep[SCS;][]{schatten71} combined with the PFSS model forms a more uniform radial field strength solution. In this way, the SCS model accounts for Ulysses's observations that revealed invariance of the radial magnetic field component with increasing latitude out of the ecliptic~\citep{wang95}. \revision{Nevertheless, some studies show that the actual source surface, better known as the Alfvén surface, is not spherical and lies anywhere from 10 to 30 solar radii above the surface of the Sun~\citep[see][]{schulz78, deforest14, cohen15}}

Plasma drags and distorts magnetic field lines and thereby distorts the coronal magnetic field from the assumed current-free configuration. Coronal models should thus account for the complex dynamics by including magnetohydrodynamical (MHD) effects and solving a set of nonlinear partial differential equations. For example, the Magnetohydrodynamics Algorithm outside a Sphere model~\citep[MAS;][]{linker99,mikic99}, the \revision{AWSoM model~\citep{vanderholst14} as part of the} Space Weather Modeling Framework~\citep[SWMF;][]{toth05}, \revision{a version of the SIP-CESE~\citep{feng10, feng20} model called CESE-HLLD~\citep{li18, li20} model}, and WindPredict-AW~\citep[][]{reville20} are such three-dimensional MHD codes. PFSS model solutions often determine their initial conditions. The MHD equations are then integrated until the plasma and magnetic fields settle into equilibrium. The solar wind plasma is constrained in closed magnetic field lines and accelerated to supersonic speeds along open field lines in the final solutions. An alternative approach situated between global MHD models and more specialized one-dimensional models is the MULTI-VP model~\citep{pinto17}. The underlying idea is to compute numerous one-dimensional model solutions that can sample sub-domains of interest and the whole solar atmosphere.

The model interface between the coronal and heliospheric domain is often determined by the topology of the coronal field extrapolation. Because the dynamic pressure term in the momentum equation is governed by the bulk speed ($\propto \rho v^2$), heliospheric solutions are sensitive to the empirically-derived solar wind speed at the model interface~\citep[see][]{riley15}. Models for specifying the speed at the interface are the Wang-Sheeley~\citep[WS;][]{wang90}, Distance from the Coronal Hole Boundary~\citep[DCHB;][]{riley01}, and the Wang-Sheeley-Arge~\citep[WSA;][]{arge03} model, updated recently with an adaptive approach~\citep[Adaptive-WSA;][]{reiss19,reiss20}. While the traditional WS model relies on an inverse relationship between the flux tube expansion rate of open field lines and solar wind speed measured at Earth, the DCHB model relies on the relation between the great circle angular distance from the nearest coronal hole boundary at the solar surface and the solar wind speed. The WSA model combines both aspects of the WS and DCHB model, and the Adaptive-WSA model continually adapts the WSA model coefficients to best match observations. \revision{Not all numerical frameworks need an empirical model to specify the physical conditions at the interface between the coronal and heliospheric domain. Frameworks such as the SWMF directly employ coronal model solutions to inform the heliospheric domain~\citep[see][]{toth12, oran13}.}

In the heliospheric domain, different approaches are used to evolve the solar wind solutions from the model interface into space. These approaches include kinematic mapping, one-dimensional upwind propagation, and global heliospheric MHD modeling~\citep{riley11b}. In kinematic mapping, the solar wind flow is treated as a chain of plasma parcels. Each parcel is accelerated or decelerated depending on the adjacent wind speed~\citep{arge00}. In contrast, heliospheric MHD models provide a complete picture of the spatial and temporal evolution of the solar wind in the inner heliosphere. Examples are HelioMAS~\citep{linker99,mikic99}, Enlil~\citep{odstrcil03}, SWMF~\citep{toth05}, SUSANOO-SW~\citep{shiota14}, LFM-helio~\citep{merkin16}, and the European Heliospheric Forecasting Information Asset~\citep[EUHFORIA;][]{pomoell18}. 

Upwind propagation tools bridge the gap between the kinematic mapping and global MHD modeling. By neglecting the pressure gradient and the gravitation term in the fluid momentum equation, upwind tools can evolve the solar wind flows in a self-consistent way. Examples are the Heliospheric Upwind Extrapolation~\citep[HUX;][]{riley11b, riley21a, issan22} and the Tunable HUX~\citep[THUX;][]{reiss20} for mapping the solar wind from the Sun to any position in the inner heliosphere, and the time-dependent HUXt~\citep[][]{owens20} model for evolving interplanetary coronal mass ejections in the ambient wind. Studying more than 40 years of data in retrospect shows that HUXt bulk speed solutions are in agreement (to within 6 percent) with MHD codes~\citep[see][]{owens20}.

\revision{Scientists have used data assimilation, a technique that "assimilates" observations (such as solar ground-based, remote-sensing and in-situ observations) into solar wind models to study the photospheric flux transport~\citep{arge10}, optimize the source surface and interface radii in coronal magnetic models~\citep{meadors20}, and predict the solar wind conditions at Earth~\citep{lang17, lang19, lang21}.} 

\revision{Furthermore, several authors have developed} useful predictive tools beyond numerical frameworks coupling the global corona and inner heliosphere. Predictive tools are often designed to forecast the solar wind at Earth while not computing the large-scale solar wind pattern. Examples are based on empirical relationships~\citep{robbins06,vrsnak07, reiss16}, machine learning algorithms~\citep{yang18, chandorkar20, bailey21}, pattern matching~\citep{bussyvirat14, riley17, owens17}, and persistence~\citep{owens13b, kohutova16, temmer18}. 

Current operational forecast services at the Met Office (Exeter, UK) and the National Oceanic and Atmospheric Administration (NOAA, Boulder, USA) rely on the coupled WSA-Enlil model framework to forecast the solar wind conditions in the inner heliosphere.

\revision{For a comprehensive introduction on coronal-heliospheric modeling, we refer the reader to~\citet{feng20}.}

\begin{figure*}
\centering
\includegraphics[width=0.99\textwidth]{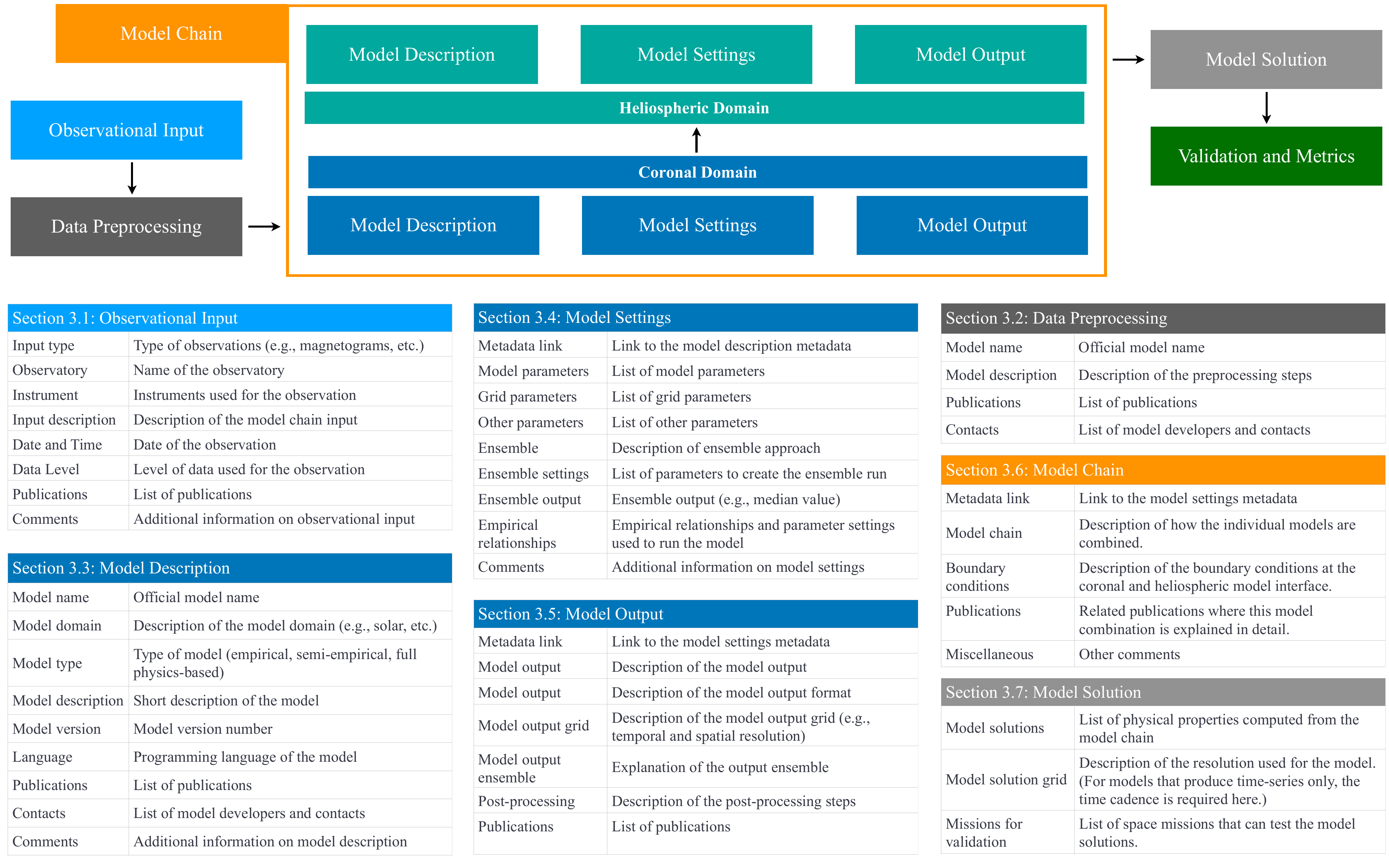}
\caption{Overview of metadata components for registering an ambient solar wind model in the database. This is an example of a standard model framework consisting of a coronal and a heliospheric model. We note that the model chain, which consists of both the coronal and heliospheric domains, is surrounded by the orange boundaries. A more detailed description of the individual components is provided in the boxes below.}
\label{fig:metadata}
\end{figure*}

\section{Metadata Architecture}\label{sec:metadata}

A comprehensive metadata architecture is needed to support a sustainable validation of solar wind models. Here we propose seven types of metadata, all of which are required to register new community models in the online platform. These metadata components include information on the observational input data, data preprocessing, model description, model setting, model output, model chain, and model solution. We will store the metadata components in an associated metadata template file according to the Space Physics Archive Search and Extract (SPASE) standards. More details on SPASE metadata can be found in~\cite{fung22_submitted}. In addition, as the backbone of the online platform, the Comprehensive Assessment of Models and Events Using Library Tools~\citep[CAMEL;][]{rastaetter19} framework hosted by NASA's CCMC will make the metadata accessible to the community online.

Figure~\ref{fig:metadata} shows the metadata architecture. The input observational data and the related data preprocessing steps are needed to document the input to the model run. Each run can consist of a chain of models used to compute the individual model solutions. The model chain template includes a description of the coronal and heliospheric model domain. As a general guideline, we distinguish between models of the corona and those of the inner heliosphere. Each model in these domains needs a detailed description of the model, settings, and outputs. It is also possible to register more than one model per domain.

Executing a model chain with different model parameters results in a new model run. The model developers will specify a model version label. The information on how the models are linked to each other is explained in the model chain metadata file. Finally, the model solution metadata describes the solution of the model chain. While also part of the metadata, the metrics provided via CCMC's CAMEL will be discussed in Section~\ref{sec:metrics}. In the following, we discuss all the components in more detail.

\subsection{Observational Metadata} 
The observational input metadata supports the reproducibility of the model solutions. Documentation of the observational input can also reduce the storage requirements for some models. This type of metadata describes the observations and explains how these observations were made. Information such as the date and time of the observation, the name of the observatory, and details on the instrumentation are required. Figure~\ref{fig:metadata} shows a list of essential information conveyed as part of the observational input metadata.

\subsection{Data Preprocessing Metadata}
The preprocessing of input data affects the output of solar wind models. Preprocessing concerns not only synoptic photospheric magnetograms in numerical frameworks but also solar imagery and other inputs to predictive tools. Ample evidence shows that deficiencies in magnetograms that serve as an inner boundary of coronal magnetic modes significantly affects the model solutions. Flux transport models such as ADAPT assimilate photospheric magnetic fields on the Earth-facing side of the Sun and apply surface flux transport to better approximate the global photospheric magnetic field. To document information on the preprocessing of the model input data, we propose the metadata information template shown in Figure~\ref{fig:metadata}.

\subsection{Model Description Metadata}
The individual models are the building blocks of the numerical framework. This type of metadata describes the individual models in the model chain used to model the solar wind in different domains. This metadata component describes each model and stores the model domain, simulation type, and version information.

\subsection{Model settings metadata}
The model settings metadata specifies the model input and settings used for computing the model solutions. In addition, it documents the model parameter settings and the data structure settings, such as grid parameters. This metadata component also includes information on expanding the model solutions from a single deterministic run to an ensemble mode.

\subsection{Model Output Metadata}
This type of metadata focuses on the description of the model output. It describes the solutions produced and explains the underlying data format and the computed physical quantities. Additionally, this metadata template file collects information about the post-processing steps after the model run that are needed to interpret the solutions.

\subsection{Model Chain Metadata}
This type of metadata describes how the individual models used to simulate the solar wind conditions from the Sun to the near-Earth space environment are connected. It focuses on how the models are linked to forecasting the solar wind at a specified location in the inner heliosphere. Furthermore, it provides details on the boundary conditions used to couple the models. 

\subsection{Model Solution Metadata}
The model solution metadata describes the final output of the model chain. It contains a list of the physical properties computed by the models and points to space missions that can be used to validate the models.

\section{Forecasting Goals and Metrics}\label{sec:metrics}

\revision{Validation goals and metrics need to cover a broad spectrum of user needs. While some users are interested in studying the large-scale heliosphere and ICME events, other users focus on the evolution of corotating interacting regions over the solar activity cycle. The operational forecasting of solar wind for modeling high-speed streams, corotating interaction regions, and ICMEs to prevent damage to spacecraft is a very different challenge. Our priority is therefore to develop a broad range of metrics that reflect the different demands in the community, in particular taking community input into account in the development of the metrics, which is one of the purposes of the ISWAT teams. We specifically calculate point-to-point metrics that are more suited for short-term studies and event-based metrics that are better suited to evaluate predictions above a certain threshold level to build some initial forecasting capabilities. While we are certainly not able to cover all the demands by the end-users, we attempt to cover some of them and hope to develop them further in the future.}

After collecting feedback from the community during workshops and conferences, we recommend testing the quality of community models in the online platform based on two forecasting goals:
\begin{enumerate}
    \item The ability of the solar wind model to forecast the temporal evolution of the solar wind speed, as well as abrupt changes from slow to fast solar wind. 
    \item The ability of the solar wind model to forecast the magnetic polarity and magnetic sector boundary crossings. 
\end{enumerate}

\begin{figure}
\centering
\includegraphics[width=0.49\textwidth]{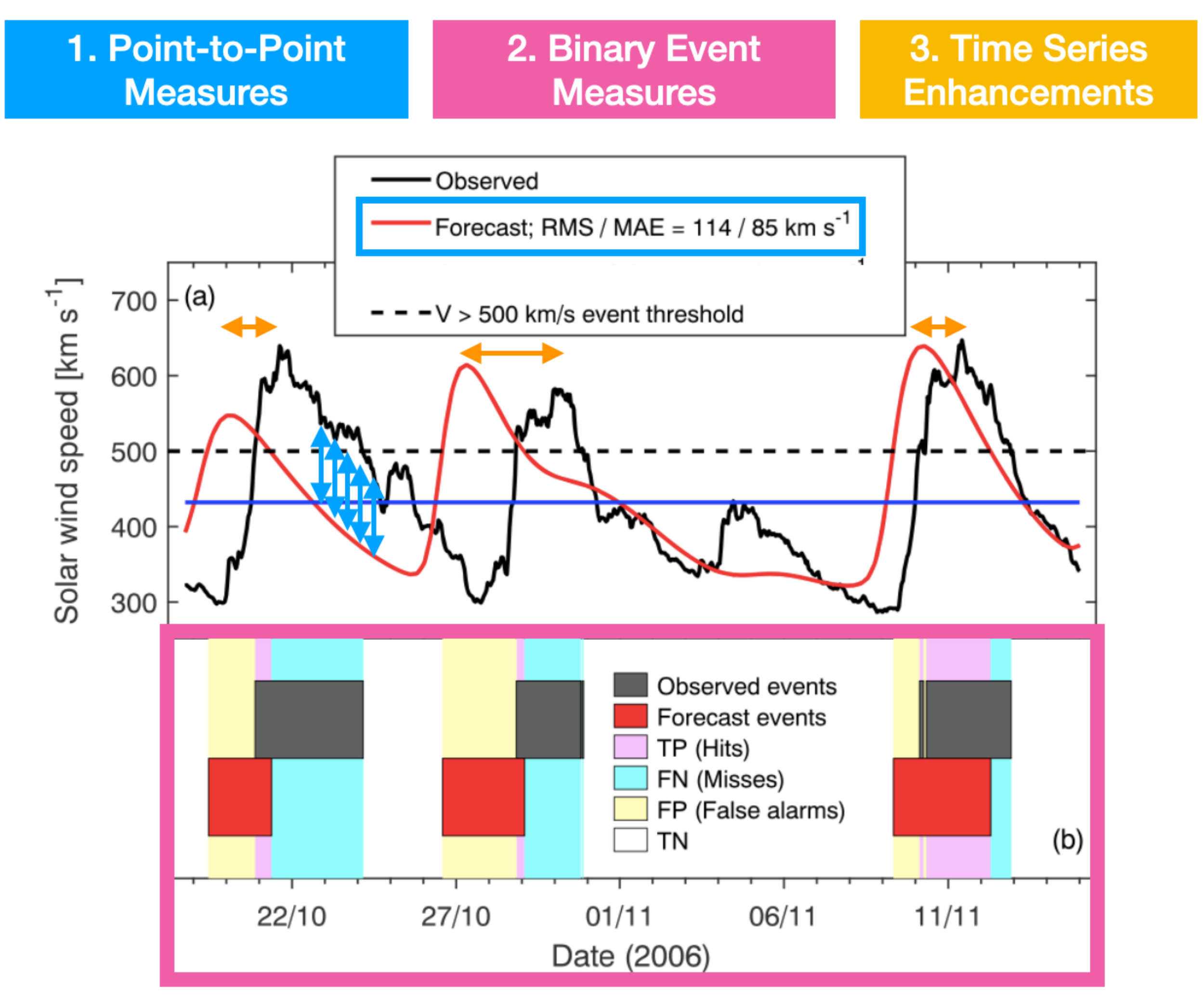}
\caption{a) Observed (black) and modelled (red) solar wind speed as a function of time with an example of proposed validation metrics explaining the concept of point-to-point measures (blue). b) binary-event measures in the pink box, and an event-based validation (orange). The solid blue line represents the climatological mean. Courtesy: Modified from~\citet{owens18}}
\label{fig:metrics}
\end{figure}

To assess the agreement between model solutions and measurements, we propose using continuous and binary variables. The first can take on any real values, while the second is restricted to categorical values such as event/non-event. Time series of solar wind properties such as bulk speed and magnetic polarity can be interpreted in both ways, as shown in Figure~\ref{fig:metrics}. First, we discuss commonly applied point-to-point comparison metrics (blue arrows). Second, we investigate the quality of model solutions in terms of binary metrics, where each time step in the predicted and observed time series is labeled as an event/non-event based on the selected threshold value (bottom panel). Third, we study procedures to detect abrupt transitions from slow to fast solar wind (orange arrows) and quantify the ability in terms of event-based metrics. 

All the validation metrics discussed here will be integrated into CAMEL's web application and made available online. In this way, end-users and model developers can use streamlined metrics and representations to compare community models. 

\subsection{Point-to-point Comparison Metrics}
We validate the quality of solar wind models by comparing their forecasts to measurements. We first focus on the underlying statistical distributions using the mean, median, and standard deviation. These measures contain valuable information on the model's tendency to over or under-estimate the observed physical properties. We moreover study the predictive skill in terms of established point-to-point error measures such as the mean error, mean absolute error, and the root mean square error. Table~\ref{tab:errorfunctions} shows the definitions of commonly used error functions, where $(f_k,o_k)$ is the $k$-th element of $n$ total forecast and observation pairs in the time series. Although strictly speaking not an error function, we also compute the Pearson correlation coefficient (PCC). 

\revision{Different point-to-point comparison metrics give insight into different aspects of the correspondence between measurement and model solution. As the difference between the mean solution and the mean measurement, the ME can be interpreted as an offset between measurement and solution. In contrast, the MSE is the squared difference between the measurement and the solution. The MAE is the arithmetic mean of the absolute difference, and represents the typical magnitude of the error. Similarly, the RMSE is the mean squared difference representing the typical magnitude of the forecast error being less sensitive to outliers.}

To complement these error measures, we use the skill score (SS) that compares the predictive skill to a simple baseline. Table~\ref{tab:errorfunctions} shows the definition of the skill score, where $\textup{MSE}_\textup{pred}$ is the mean square error of the forecasted time series, and $\textup{MSE}_\textup{ref}$ is the MSE of a reference baseline model. A common example of a baseline model in the literature is the climatological mean defined as the mean value of the observation~\citep{owens18}. For an ideal forecast, the MSE results in zero, and the SS consequently results in a value of 1. A forecast that equals the skill of the climatological mean results in an SS value of 0, and a forecast that is less skillful than the baseline model results in a negative SS value. Other reference models, such as a model of 27-day persistence (assuming that the solar wind conditions will repeat after each Carrington rotation), can replace the climatological mean. 

\begin{table}[]
\caption{Overview of point-to-point comparison metrics.}
\begin{center}
\begin{tabular}{lll}
\hline
Metric                   & Short Name & Definition \\ \hline
Mean error             & ME         & $\frac{1}{n} \sum_{k=1}^n (f_k - o_k)$    \\ \\
Mean square error      & MSE        & $\frac{1}{n} \sum_{k=1}^n (f_k - o_k)^2$           \\ \\
Mean absolute error    & MAE        & $\frac{1}{n} \sum_{k=1}^n \left|f_k - o_k\right|$           \\ \\
Root mean square error & RMSE       & $\sqrt{\frac{1}{n} \sum_{k=1}^n (f_k - o_k)^2}$           \\ \\
Skill score            & SS         & $1 - \frac{\rm{MSE}}{\rm{MSE}_{\rm{ref}}}$           \\ \hline
\end{tabular}%
\end{center}
\label{tab:errorfunctions}
\end{table}

\subsection{Binary Metrics}
The previous section quantified the magnitude of the error at every time step. This section focuses on an alternative procedure that classifies each time step as an event/non-event. \citet{owens18} summarized the advantages of this procedure as follows. First, traditional error functions do not differentiate between times of slow and fast solar wind conditions. Forecasters, however, are often only interested in times when the solar wind conditions exceed a certain threshold level, while the temporal evolution of the slow solar wind is secondary. Second, outliers in the forecasted time series can significantly affect the computed error functions and correlation coefficients. For these reasons, a reasonable approach is to classify each time step in the solar wind model solutions and measurements as an event/non-event. 

To define events and non-events in solar wind time series, we first need to define an event threshold. By cross-checking events and non-events in the predicted and observed time series, we can compute the number of hits (true positives; TPs), false alarms (false positives; FPs), misses (false negatives; FNs), and correct rejections (true negatives; TNs). TPs are correctly forecasted events, while FNs are observed events that were not forecasted. Similarly, FPs are forecasted events that were not observed, and TNs are correctly forecasted non-events. We can compute a set of skill measures from the total counts of the different scenarios, summarized in the so-called contingency table. Table~\ref{tab:binarymetrics} shows some examples. 

A noteworthy skill measure is the so-called true skill statistics (TSS). The TSS is defined in the range $[-1,1]$. While a TSS of 0 indicates no skill, a perfect model solution would score 1 (or -1 for an inverse forecast). The TSS has the advantage that it is unbiases by the ratio between forecasted and observed events and uses all entries in the contingency table~\citep{bloomfield12}. 

\subsection{Event-based Validation}
Relying on point-to-point comparison metrics can be misleading~\citep[see, for example,][]{owens05, macneice09a, macneice09b}. This is the case when the temporal evolution of solar wind properties is generally well-predicted, but the arrival times differ in forecast and observation. \cite{owens05, macneice09b, reiss16} proposed a three-stage strategy to quantify these arrival time errors in bulk speed forecasts. First, is the definition and detection of abrupt transitions from slow to fast solar wind speed in forecast and measurements. Second is the association of the events detected in the forecasts and observations. Third, is the computation of skill measures to compare the predictive abilities of the models investigated. Table~\ref{tab:binarymetrics} shows several validation measures. 

\revision{We specifically focus on abrupt changes from slow to fast solar wind for the event-based validation. In the first version of the online platform, we will define these events according to the criteria discussed in~\cite{reiss16}. In a later version of the platform, we will work on algorithms to automatically detect and assess magnetic sector boundary crossings events.}

\begin{table}[]
\caption{Overview of binary metrics defined by the entries of a contingency table.}
\begin{center}
\begin{tabular}{lll}
\hline
Metric                 & Short Name & Definition \\ \hline
True Positive Rate     & TPR        & $\frac{\rm{TP}}{\rm{TP} + \rm{FN}}$    \\ \\
False Positive Rate    & FPR        & $\frac{\rm{FP}}{\rm{FP} + \rm{TN}}$           \\ \\
Threat Score           & TS         & $\frac{\rm{TP}}{\rm{TP} + \rm{FP} + \rm{FN}}$           \\ \\
True Skill Statistics  & TSS        & $\rm{TPR} - \rm{FPR}$           \\ \\
Bias                   & BS         & $\frac{\rm{TP} + \rm{FP}}{\rm{TP} + \rm{FN}}$           \\ \hline
\end{tabular}%
\end{center}
\label{tab:binarymetrics}
\end{table}

\section{An Open Platform for Model Validation} \label{sec:platform}
We use the existing Comprehensive Assessment of Models and Events Using Library Tools~\citep[CAMEL;][]{rastaetter19} framework to make our developments accessible to the community. The implementation of CAMEL consists of a front-end and a back-end reaching deep into CCMC's modeling infrastructure. 

The front-end of CAMEL will make our results visible to the community. It will enable end-users to visualize state-of-the-art solar wind solutions and compare them with spacecraft measurements and each other. By selecting time intervals of interest, end-users will be able to compute the validation metrics discussed in Section~\ref{sec:metrics}. The repertoire of metrics will be expanded and updated in later versions of the platform to stay up-to-date with operational and research user needs. CAMEL provides data interpolation options such as nearest neighbor and linear interpolation for the computation of these metrics. CAMEL's front-end will furthermore make all the metadata information discussed in Section~\ref{sec:metadata} publically available. By doing so, end-users will be able to access all the background information needed to interpret the model solutions displayed in the online platform. 

The back-end of CAMEL links to existing services at the CCMC. This includes, for example, an Application Programming Interface (API) allowing end-users to download the model solutions and computed metrics displayed in the CAMEL front-end. In a future step, we will offer solar wind model developers the opportunity to feed the solutions of models currently installed at the CCMC directly into CAMEL. This way, model developers can make their models and updated versions, including a comprehensive validation analysis, readily accessible to the community. For more information on CAMEL, we refer the interested reader to~\cite{rastaetter19}.

\section{Discussion and Outlook} \label{sec:discussion}
Our ability to model the global corona and solar wind has grown significantly over the past decades. The validation of state-of-the-art model solutions has, however, stagnated. Validation of the quality of models for selected events and time intervals, usage of individually developed metrics, and a slow iterative process between developers and end-users cause a bottleneck. These validation practices make a complete assessment of the state-of-the-art difficult or even impossible. This paper has discussed the first steps of the Ambient Solar Wind Validation Team to address these challenges by streamlining the validation of solar wind models. By collecting feedback during team meetings and international conferences, our team has identified forecasting goals of importance to the research and forecasting communities, developed a metadata architecture to register solar wind models, and recommended metrics that reflect a broad spectrum of scientific and forecasting needs. We have also taken the first steps towards developing an open online platform by feeding all these developments into CAMEL~\citep{rastaetter19} hosted by NASA's CCMC. 

The online platform proposed here does not solve the open problems in large-scale solar wind modeling, but it does help to bridge the gap between developers and end-users and thereby accelerate the feedback loop to drive innovation and progress in solar wind model development. Our approach for streamlining the validation in space weather research is not new; the need for automated validation protocols has been identified before in~\cite[][]{macneice18b}, and similar community strategies have been successfully implemented in adjacent space weather disciplines. The \href{https://www.iswat-cospar.org/assessment}{International Forum for Space Weather Capabilities Assessment} is devoted to creating a community-wide consensus on continuous assessment of space weather predictive capabilities in line with the COSPAR Space Weather Roadmap~\citep{schrijver15}. Participating teams in this forum such as the \href{https://iswat-cospar.org/H2-01}{CME Arrival Time and Impact Working Team} and the \href{https://iswat-cospar.org/H3-01}{Solar Energetic Particle Validation Team} have developed a similar approach for assessing the predictive skill of models~\citep[see][]{verbeke19}. Other examples are \cite{pulkkinen13}, \cite{glocer16}, and the GEM CEDAR Challenge.

While our effort can provide a step forward in validating the state-of-the-art, additional factors need to be considered. First, we expect the proposed metadata architecture to capture most of the details required to make the model solutions reproducible. However, we cannot claim that the metadata will completely capture all of these details. Frameworks of coupled corona and inner heliosphere models are sophisticated numerical algorithms with many parameters and settings. Second, the online platform is designed to capture the current state-of-the-art. Although considerably faster than reporting new model results in the literature, our approach in the present form still needs overhead to upload new solutions manually. Third, and related to the last note, in an ideal case, the validation should be carried out by an independent, unbiased agent, without model developers uploading individual solutions to the online platform. In the future, we expect to use model runs directly from the CCMC, with the model developer's permission, to allow solutions to be fed directly into the online platform. Many models are already hosted at the CCMC. We argue that trust in the scientific integrity of participating models is necessary, and the opportunity to upload model solutions from external sources will continue to allow a comprehensive comparison of community models.

In the near future, work is ongoing to implement and improve upon these first steps. Most metrics have already been implemented in a beta version of the online platform. In the next step, we will release the first version of the platform online to the public, including the complete list of metrics and access to the metadata information of participating models. Specifically, we will explore the physical properties of the evolving solar wind flow such as solar wind speed, density, temperature, magnetic polarity, and magnetic sector boundary crossings with either point-to-point comparison metrics and more advanced event-based validation metrics. We will also continue to advertise the open platform in the space weather community to add more solar wind models and provide a more comprehensive state-of-the-art validation. Almost all model developers listed in Figure~\ref{fig:stateoftheart} have agreed to provide their model solutions to the online platform. Furthermore, we will focus on new data available from recently launched space missions such as Parker Solar Probe~\citep[][]{fox16} and Solar Orbiter~\citep[][]{muller13}, and expand our repertoire of metrics to validate probabilistic model solutions. 

In our long-term planning, one future avenue is to use the infrastructure developed here to implement an Ambient Solar Wind Scoreboard to assess predictive model capabilities in real-time. A real-time assessment and display of large-scale solar wind models can be a valuable tool for improved space weather awareness at Earth and other planetary environments.

\section{Summary} \label{sec:summary}
The \href{https://www.iswat-cospar.org/h1-01}{Ambient Solar Wind Validation Team} embedded in the Committee on Space Research (COSPAR) - \href{https://www.iswat-cospar.org}{International Space Weather Action Teams (ISWAT)} initiative was formed to support the space weather community to maintain up-to-date solar wind model validation and track progress over time. This study presents our first steps towards developing an open platform for validating solar wind forecasting models. We propose a metadata architecture consisting of seven components to support a continuous, transparent, and reproducible validation of solar wind models. Next, we recommend community-wide forecasting goals, including the forecast of the temporal evolution of the solar wind and abrupt changes from slow to fast solar wind, as well as the forecast of the magnetic polarity and magnetic boundary crossings. We furthermore discuss a comprehensive validation procedure based on the point-to-point comparison, binary, and event-based metrics. We conclude with a status update of the online platform and outline future perspectives of this community effort. 

\section{Acknowledgments}
The authors thank all contributors to the COSPAR ISWAT initiative, which supported this research effort. The authors acknowledge the following organizations and programs: M.A.R., and C.M., acknowledge the Austrian Science Fund (FWF): P34437, J4160-N27, P31659, P31521; K.M.~acknowledges support by the NASA HGI program (\# 80HQTR19T0028), the NASA cooperative agreement NNG11PL10A and 80NSSC21M0180; J.J.G.A.~acknowledges to Investigadores por M\'exico-CONACYT (CONACYT Fellow), CONACYT LN 315829 (2021) and CONACYT-AEM 2017-01-292684 grants. S.D. acknowledges support from the argentianean grants PICT-2019-02754 (FONCyT-ANPCyT) and UBACyT-20020190100247BA (UBA). \revision{The authors thank two anonymous reviewers and the editor for their constructive criticism that improved the presentation of our results and made the content more appealing to a broader audience.}

\bibliographystyle{model5-names}
\biboptions{authoryear}
\bibliography{main}

\end{document}